\documentclass[runningheads]{llncs}
\usepackage[T1]{fontenc}
\usepackage{listings} 
\usepackage{xcolor}
\usepackage{lipsum} 
\usepackage{url} 
\usepackage{graphicx}
\usepackage{hyperref}
\begin{document}
\title{Object-Centric Analysis of XES Event Logs: \\
Integrating OCED Modeling with SPARQL Queries}
\titlerunning{Object-Centric Analysis of XES Event Logs}
%
\author{Saba Latif\inst{1},
Huma Latif\inst{2} \and
Muhammad Rameez Ur Rahman\inst{3}}
\authorrunning{L. Saba et al.}
%
\institute{Sapienza University of Rome, Italy \\ \email{saba.latif@uniroma1.it} \and
University of Sahiwal, Pakistan \\ \email{latifhuma666@gmail.com} \and
Ca' Foscari University of Venice, Italy\\
\email{muhammad.rahman@unive.it}}

\maketitle            
\begin{abstract}
Object Centric Event Data (\textsc{OCED}) has gained attention in recent years within the field of process mining. However, there are still many challenges, such as connecting the XES format to object-centric approaches to enable more insightful analysis. It is important for a process miner to understand the insights and dependencies of events in the event log to see what is going on in our processes. In previous standards, the dependencies of event logs are only used to show events, but not their dependencies among each other and actions in detail as described in \textsc{OCEDO}. There is more information in the event log when it is revealed using the \textsc{OCEDO} model. It becomes more understandable and easier to grasp the concepts and deal with the processes. This paper proposes the use of Object-Centric Event Data Ontology (OCEDO) to overcome the limitations of the XES standard in event logs for process mining. We demonstrate how the \textsc{OCEDO} approach, integrated with SPARQL queries, can be applied to the BPIC 2013 dataset to make the relationships between events and objects more explicit. It describes dealing with the meta descriptions of the \textsc{OCEDO} model on a business process challenge as an event log. It improves the completeness and readability of process data, suggesting that object-centric modeling allows for richer analyses than traditional approaches. 

\end{abstract}
\section{Introduction}
The XES standard\footnote{https://www.tf-pm.org/resources/xes-standard/about-xes} is an XML-based format for traditional event logs. The XES standard describes event logs where a single case notion, that is, a process instance, needs to be chosen, and it is part of the Task Force on Process Mining (TFPM) \footnote{https://www.tf-pm.org/upload/1678694478319.pdf}. However, in real life, information systems like SAP ERP systems support processes that cannot be converted into a single case notation. Many objects interact with each other, and these interactions overlap. A single uniform notation to represent this behavior is missing. This leads to major data loss and prevents the actual information about object dependencies from being shown, which may cause convergence or divergence problems and result in incomplete process models \cite{Ghahfarokhi2023MultiDimensional}. 

\begin{figure}[t!]
\includegraphics[width=\textwidth]{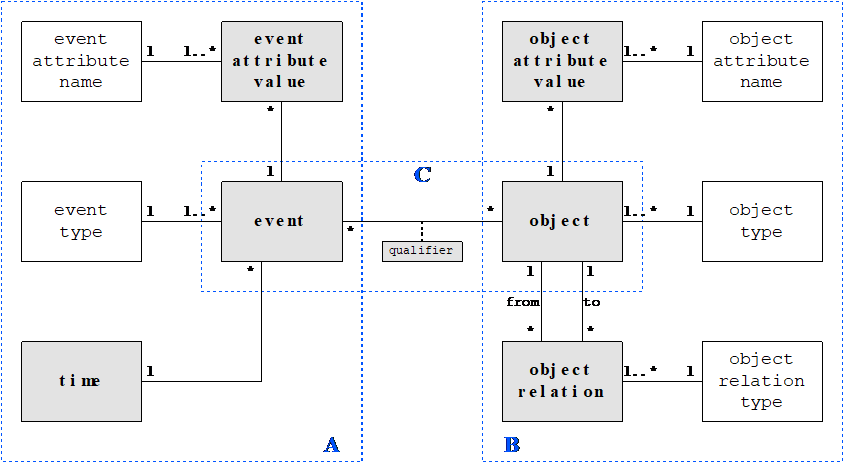}
\caption{The visualization of Object Centric Event Data(OCED) meta model} \label{fig2}
\end{figure}

Object centric process mining is a new paradigm which focuses on the object centric event logs. There are many researchers exploring this paradigm as in \cite{Miri2025OCPM2}. 
The Object Centric Event Data (OCED) standard can handle processes in an object-centric way by allowing multiple notations and extracting the dependencies between them as in \cite{9980730}. These types of logs are not flat logging formats as in XES. The purpose of this paper is to provide a general standard for Object-Centric Event Data ontology(OCEDO)\footnote{https://www.tf-pm.org/upload/1700818140843.pdf} and to show its working using a GraphDB schema to solve BPIC 2013 challenge question as an example. This is a new perspective towards solving BPIC 2013 questions using a novel meta model in an object-centric way\cite{berti2021object} using semantics. It is similar to the Object-Centric Event Logs (OCELs) \cite{Berti2024OCEL2}\footnote{https://ocel-standard.org/}, but while OCEL deals only with the logs, this approach deals with the data and their dependencies. Some logging formats have been defined to address this problem, as discussed before, but they were not widely used due to their complex structures and performance issues, for example, XOC allows reconstructing the entire database state \cite{berti2023oc}.
This standard aims to provide a generic way to change event data using multiple case notations for exchanging data between process mining tools and information systems.\\
When developing the OCEDO standard \footnote{https://www.tf-pm.org/resources/oced-standard}, the goals were interoperability to handle many languages and enable understanding across systems and platforms, generalization to support storing and using events, objects, and their attributes with possible extensions, provision of a collection of examples related to event logs and supporting information from information systems, and tool or library support for implementation in custom applications like \cite{Goossens2022Enhancing}, \cite{Berti2024OCEL2}. 
The advantages of using OCEDO are that it incorporates object-centric notions using ontology and has been extensively discussed within the working group. Simplicity compared to expressiveness has been considered so that the standard is easy to understand and grasp. Standardization versus adoption has also been considered, with a focus on ensuring conversion steps are minimal and clear using knowledge graphs as in \cite{Swevels2024ImplementingOCED}. The OCED meta-model has already been circulated to the community, and feedback was sought in 2022. By structuring event data through an object-centric lens, the OCED standard aims to make process data clearer, more complete, and better suited to modern, complex information systems. The meta-model has already been shared with the community, and feedback was gathered in 2022 to refine its design and usability as shown in Figure \ref{fig2} taken from \cite{fahland2024towards}.
This paper explores how Object-Centric Event Data Ontology(OCEDO) can enrich process mining, focusing on transforming the BPIC 2013 dataset into an object-centric representation. We argue that OCEDO captures dependencies and relationships that traditional XES-based event logs often miss. We describe the OCEDO meta-model, compare it to XES and OCEL, and demonstrate how event-object relationships can be queried using SPARQL to gain deeper insights. The main purpose of this work is to illustrate how OCEDO can be used to convert XES logs into an object-centric format, providing additional insights and more meaningful results. We address the technical aspects of process mining and event log modeling in an object-centric manner. The paper offers a concrete example of how OCEDO structures can be represented and queried using the BPIC 2013 event log and solving one question asked in BPIC 2013 event log. Additionally, OCEDO complements existing standards, such as OCEL 2.0. Its purpose is to work alongside these tools, offering an alternative or supplementary option based on the user's requirements.

In Section 2, we present the BPIC 2013 Challenge Event Log description and
comparison of OCEDO with other approaches. Section 3 describes the explanation of the Event-Object Relationship of OCEDO in SPARQL Query. Section 4 discusses related work on existing standards and techniques for modeling event logs and object-centric data, and finally, Section 5 concludes the paper and outlines directions for future research.
\begin{figure}[b!]
    \centering
    \includegraphics[width=1\linewidth]{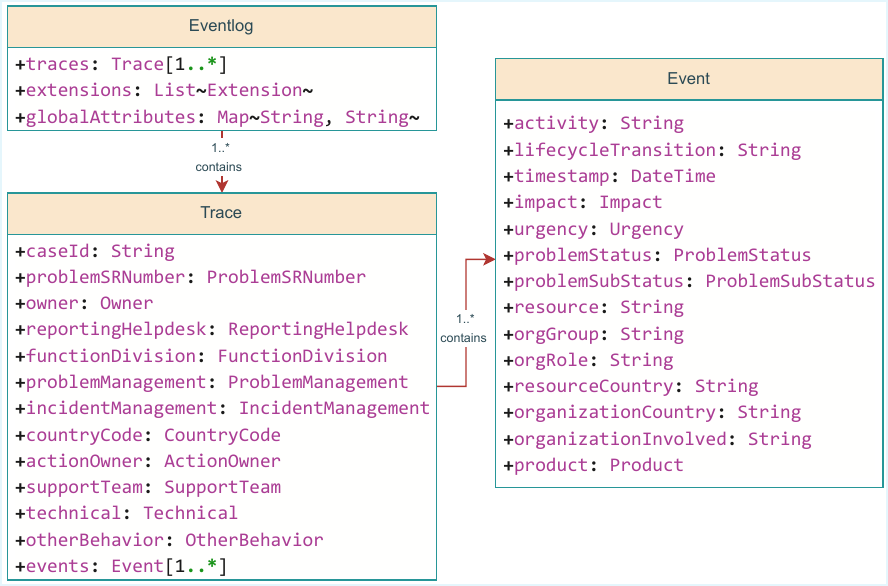}
    \caption{The abstract class diagram of the BPIC 2013 event log in XES representing event log, trace, and event attributes. The full diagram is available at the link \url{https://tinyurl.com/BPIC2013ClassDiagram}}.
    \label{fig:classdiagram}
\end{figure}

\section{BPIC 2013 Challenge Event Log}
The BPIC 2013 event log\cite{vanDongen2013BPIC2013} is a real-life dataset often used in process mining research. It contains events related to the customer service process of a large Dutch insurance company \cite{vanderAalst2016}. Each event in the log records activities like receiving, handling, and closing customer requests in process mining \cite{vanderAalst2015}. 
This event log is valuable because it shows how cases move through different departments and how long each step takes. It includes details such as the timestamp of each event, the resources involved (like employees or teams), and the type of service requested. Researchers and analysts use the BPIC 2013 log to discover patterns, find inefficiencies \cite{carmona2018}, and suggest improvements in business processes \cite{deLeoni2015}. 
It helps in understanding what work happens and how it is compared to how it is supposed to happen. 
A class diagram of the whole BPIC 2013 challenge in XES format is shown in Fig. \ref{fig:classdiagram}.
By analyzing this event log, we can gain insights into real operational challenges and performance issues in service processes. An example of the BPIC 2013 challenge in the OCED meta model representation is shown in Figure \ref{fig:classdiagram}. We did this conversion by using the techniques of representing event logs in OCEDO model using semantics. {A diagram showing BPIC 2013 event log example converted in OCEDO representation where events and objects are related with qualifiers and object to object relations \ref{fig3}. The final output of our technique is in turtle file format, which can be used as it is in SPARQL queries analysis. The representation is depicted in a turtle file in the OECD model.

\begin{figure}[t!]
\includegraphics[width=1.0\textwidth]{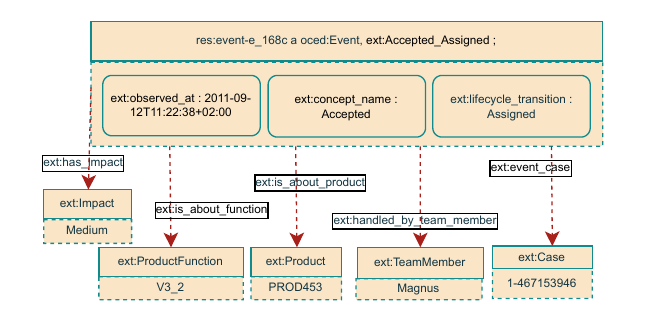}
\caption{An example showing BPIC 2013 event log example converted in OCEDO representation where events and objects are related with qualifiers and object to object relations in object centric manner} \label{fig3}
\end{figure}

\section{Explanation of the Event-Object Relationship of OCEDO into SPARQL Query}
The demonstration of BPIC 2013 challenge into OECD is derived from the repository link \footnote{https://gitlab.isis.tuwien.ac.at/Ekaputra/ocedo}. Based on the information provided here, we are analyzing the BPIC 2013 event log. This section elaborates on the purpose, methodology, and advantages of the provided SPARQL query, specifically designed to interact with data structured according to the file named 2013 small integrated ttl.
Based on the provided information, we analyze the BPIC 2013 event log and answer one of its questions to demonstrate the power of our approach.\footnote{\url{https://ceur-ws.org/Vol-1052/vinst\_data\_set.pdf}}
The core question addressed by this query revolves around understanding the intricate relationships between events and objects within a given dataset. It seeks to uncover specific instances where an event is associated with an object, along with their respective classifications, types, and any relevant descriptive attributes. Fundamentally, it aims to provide a comprehensive, granular view of the event-object landscape as modeled by the OCED ontology(OCEDO), allowing for detailed data exploration.
This question is precisely solved by leveraging the explicit ext:EventObject class and its foundational properties defined within the 2013 small integrated turtle file. The query systematically identifies every ext:EventObject instance, connecting an event (ext:event) to an object (ext:object). For each such connection, it proceeds to retrieve auxiliary information: an optional classifier value that might categorize the event object pairing; the event type and time for the associated event; and the object type for the related object. Crucially, the strategic use of OPTIONAL clauses ensures that the query remains robust, returning results even if certain attributes or properties are not present for every single event or object, thus preventing the exclusion of valuable core relationships. By performing a join across these related entities and their properties, the query constructs a cohesive record for each event-object linkage.
The benefits derived from employing this query are multifaceted. Firstly, it enables thorough data exploration and validation of the event-object model, confirming that relationships and attributes are correctly represented as per the ontology's design. Secondly, it provides a foundational dataset for qualitative analysis, allowing users to identify specific event-object interactions, understand their context through types and classifiers, and examine granular details via attributes. This is particularly valuable for scenarios where understanding individual event impacts on specific objects, or vice versa, is paramount. Thirdly, by exposing all relevant properties in a structured output, the query facilitates further data processing and visualization. The results can be easily consumed by other tools or scripts for more advanced analytics, reporting, or graphical representations, aiding in pattern recognition and anomaly detection within event-object dynamics. Lastly, its adherence to the ontology's structure makes it a highly interpretable and maintainable query, directly reflecting the logical design of the knowledge graph as shown in figure \ref{fig:queryoutput}.
\subsection{Questions of BPIC 2013 challenge}
The process owner poses a number of possible questions on which answers are sought:
\begin{enumerate}
    \item \textbf{Push to Front (incidents only)}: Is there evidence that cases are pushed to the 2nd and 3rd line too often or too soon?
    \item \textbf{Ping Pong Behavior}: How often do cases ping pong between teams, and which teams are more or less involved in ping-ponging?
    \item \textbf{Wait User Abuse}: Is the "wait user" substatus abused to hide problems with the total resolution time?
    \item \textbf{Process Conformity per Organization}: Where do the two IT organizations differ, and why?
\end{enumerate}

\subsubsection{Ping-Pong Behavior}
This analysis addresses Question 2 from the BPIC 2013 challenge, focusing on identifying ping-pong behavior between support teams. The objective is to quantify how frequently cases bounce between teams and determine which teams are most involved in this inefficient practice. Our solution implements a sophisticated pattern-matching approach that detects when a support team handles a case, transfers it to another team, and subsequently receives the case back—indicating workflow inefficiencies and potential process gaps. The analysis identifies the specific teams engaged in ping-pong exchanges, measures the frequency of such occurrences, and ranks teams by their level of involvement. By providing clear metrics on inter-team case transfers, this analysis enables organizations to pinpoint problematic handover patterns and implement targeted process improvements to reduce unnecessary case bouncing and improve overall support efficiency.

\subsubsection{Ping-Pong Behavior Analysis}
This section elaborates in detail on the purpose, methodology, and advantages of the provided SPARQL query, which is specifically designed to effectively interact with data structured according to the \texttt{2013\_small-integrated.ttl} ontology. The core question addressed by this query revolves around understanding the intricate relationships between events and objects within a given dataset, with the explicit goal of identifying and quantifying the impact of the \textbf{'Ping-Pong behavior'} event on \textbf{accepted cases}. The analysis addresses the problem of detecting ping-pong behavior in support incident handling by implementing a sophisticated pattern matching approach. This solution identifies cases where a support team initially handles a case, transfers it to another team, and subsequently receives the case back a pattern that indicates inefficient workflow and potential process gaps. The core innovation lies in using an EXISTS subquery that searches for a specific three-event sequence: Team A handles the case at time T1, Team B handles it at a later time T2, and Team A again handles it at an even later time T3, with the critical constraint that Team A and Team B must be distinct entities. This approach successfully solves the detection problem by efficiently scanning event sequences without requiring consecutive event matching, thereby capturing ping-pong behavior even when intermediate handling events occur between the key transitions. The implementation aggregates results by case while calculating temporal boundaries, providing a comprehensive view of each case's lifecycle alongside a clear boolean indicator of ping-pong occurrence, enabling effective monitoring and process optimization.

\begin{lstlisting}[language=SPARQL]
PREFIX ocedo: <https://w3id.org/ocedo/core#>
PREFIX ext: <https://w3id.org/ocedo/ext#>
PREFIX xsd: <http://www.w3.org/2001/XMLSchema#>

SELECT ?case
       (COUNT(?pingPong) > 0 AS ?hasPingPong)
       (MIN(?time) AS ?minTime)
       (MAX(?time) AS ?maxTime)
WHERE {
  {
    SELECT ?case ?time (GROUP_CONCAT(?team) AS ?teams)
    WHERE {
      ?event ext:event_case ?case ;
             ocedo:observed_at ?time ;
             ext:handled_by_support_team ?team .
    }
    GROUP BY ?case ?time
    ORDER BY ?case ?time
  }
  BIND(
    EXISTS {
      ?e1 ext:event_case ?case ;
          ext:handled_by_support_team ?teamA ;
          ocedo:observed_at ?time1 .
      ?e2 ext:event_case ?case ;
          ext:handled_by_support_team ?teamB ;
          ocedo:observed_at ?time2 .
      ?e3 ext:event_case ?case ;
          ext:handled_by_support_team ?teamA ;
          ocedo:observed_at ?time3 .
      FILTER(?teamA != ?teamB && ?time1 < ?time2 && ?time2 < ?time3)
    } AS ?pingPong
  )
}
GROUP BY ?case
ORDER BY ?hasPingPong
\end{lstlisting}

\begin{figure}
    \centering
    \includegraphics[width=1\linewidth]{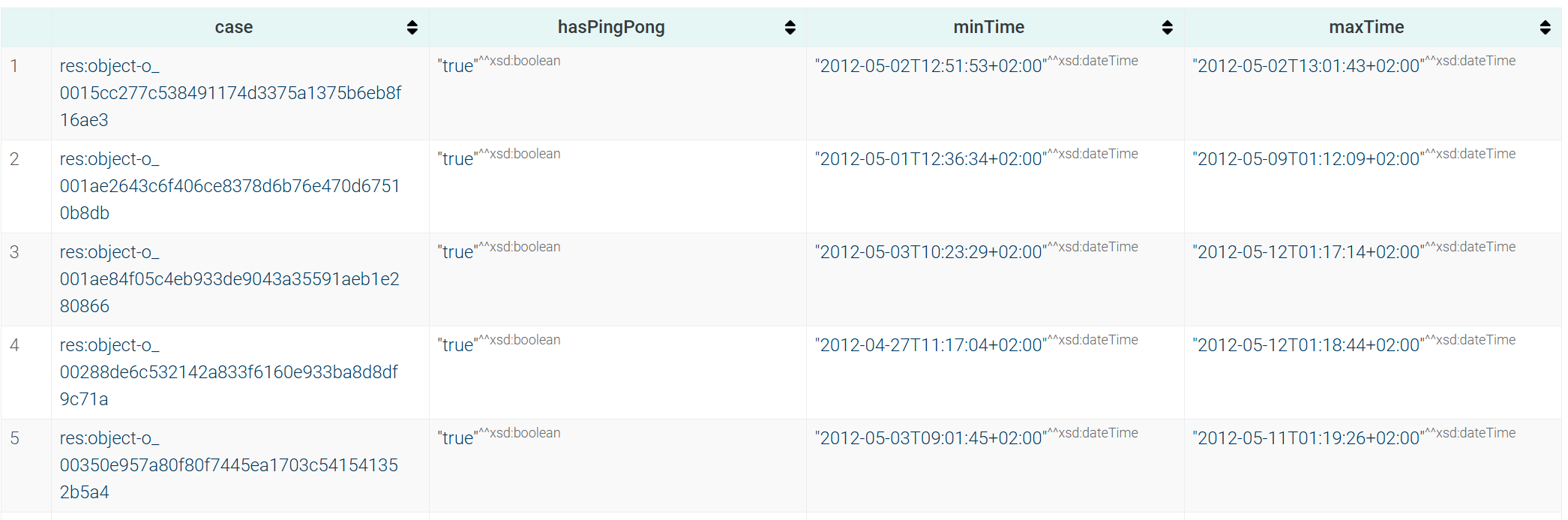}
    \caption{Result of BPIC 2013 in the form of a Graph in GraphDB extracted by running the above SPARQL Query}
    \label{fig:queryoutput}
\end{figure}
\section{Related Work}
\label{sec:related-work}
The field of \emph{object-centric process mining} (OCPM) has emerged as a significant advancement beyond traditional event-centric approaches, enabling more comprehensive analysis of business processes involving multiple interacting objects. Also, event logs are the basis for applying process mining techniques. A typical event log consists of a case ID, which represents a unique instance of a process \cite{berti2021object}, for example, a student or any product available for sale online. It includes other tags such as an activity, a timestamp, and additional characteristics in the form of key-value pairs, such as login, booking a test, or placing an online order in the form of XES event logs. Further, the study focusing on how event logs are represented, as they are essential for process mining by recording the sequence, context, and relationships of activities, is presented in \cite{G16}.  
\begin{table}[b!]
\label{TAB:1}
\caption{Comparison of OCEDO approach with XES and OCEL event logs providing difference and similarities}\label{tab1}
\begin{tabular}{|l|p{3.1cm}|p{3.1cm}|p{3.1cm}|}
\hline
Key Aspect & $\textsc{OCEDO}$\footnote{https://www.tf-pm.org/resources/oced-standard} & XES\footnote{https://www.tf-pm.org/resources/xes-standard} & OCEL\footnote{https://ocel-standard.org/} \\
\hline
Flexibility & Focus on events and objects where object interactions trigger events & Stream of data as independent events in a generic, loosely-coupled manner & Object-centric, allowing events to relate to multiple objects, improving realism \\
\hline
Extensibility & Rigid standardized approach with predefined object attributes and behavior & Easier adaptation to changing requirements & Moderate supports extensions but still constrained by JSON/XML schema \\
\hline
Scalability & Challenges due to tightly coupled objects and complex events & Stream-based, more scalable with large volumes and parallel processing & Better scalability than XES by grouping events with multiple objects, though still bound by storage overhead \\
\hline
Interoperability & Generic knowledge graph format easily shared across systems & Tighter dependency on XML limits interoperability & Designed for interoperability, using JSON and XML formats widely supported \\
\hline
Event Processing & Event driven programming where events trigger specific object actions & Requires parsing of flat event streams; events not tied to objects & Supports querying across objects and events, better suited for complex analysis \\
\hline
\end{tabular}
\end{table}
Recent research has made substantial contributions to this paradigm through theoretical foundations, practical frameworks, and novel applications. The conceptual foundation was established through the formalization of \emph{object-centric event logs} (OCEL), with \cite{Berti2024OCEL2} introducing the OCEL 2.0 specification that standardizes representation formats. Building on this, \cite{9980730} provided crucial definitions for cases and variants in object-centric contexts, while \cite{berti2023oc} developed the OC-PM framework for analyzing both event logs and process models from an object-centric perspective.
Several extensions have enhanced the capabilities of object-centric approaches. \cite{Goossens2022Enhancing} proposed methods for improving data-awareness in object-centric logs, and \cite{Miri2025OCPM2} introduced OCPM\textsuperscript{2} to address event data extraction challenges. The practical implementation aspects were explored by \cite{Swevels2024ImplementingOCED} through knowledge graph representations. The examination of formal methods establishes essential groundwork for enhancing process mining approaches, especially within object-centric paradigms. Further, techniques from formal methods offer substantial methodological support that can strengthen process mining frameworks. TLA+ has proven effective for specifying and verifying concurrent processes in distributed systems \cite{Latif2019Waste}, while finite automata provide systematic structures for modeling state-based control logic and checking event sequence compliance \cite{Rehman2019Railway}. Graph based modeling demonstrates significant utility in representing complex relationships in resource allocation and state transition scenarios \cite{Latif2018Parking}. The application of formal verification methods to ensure smart contract robustness in blockchain contexts \cite{Latif2025Formal} indicates a viable approach for managing object-centric event log complexities. Additionally, formal taxonomies enable organized classification of security threats and process deviations \cite{Latif2017IoT}, and complex event processing techniques facilitate real-time monitoring and analysis of event streams \cite{Latif2018Traffic}. These established methodologies constitute the foundational basis for OCEDO, representing a new paradigm that integrates formal verification rigor with object-centric process mining principles.The BPIC 2013 dataset \cite{vanDongen2013BPIC2013} has served as an important benchmark, with earlier process mining techniques \cite{vanderAalst2016,vanderAalst2015,deLeoni2015} laying the groundwork for multi perspective analysis that later influenced object-centric approaches. Comparative studies like \cite{carmona2018} demonstrated the advantages of richer log representations for variant analysis. Recent systematic evaluations by \cite{Berti2023LiteratureReview} have documented the field's progress while identifying remaining challenges. Emerging applications are expanding the paradigm's reach, including innovative work by \cite{Hobeck2025BlockchainOCEL} applying object-centric methods to blockchain transaction analysis. While these advances have established OCPM as a powerful paradigm, open challenges remain in areas of scalability, standardization, and integration with existing process mining infrastructures. In this table \ref{tab1}, we are showing a comparison of OCEDO with other approaches to show its viability. We are presenting a comparison with XES, OCEL and OCEDO techniques from different perspectives including flexibility, extensibility, scalability, interoperability and event processing.
\section{Conclusion}
This paper explored the application of Object-Centric Event Data in the context of business process analysis, particularly in addressing the BPIC 2013 challenge questions. We demonstrated how event logs can be enriched to reveal deeper insights into process dependencies, interactions, and underlying behaviors going beyond traditional event-centric approaches that often overlook intricate relationships between events and objects. The study highlighted the advantages of OCEDO in enhancing process mining by providing a more comprehensive, structured, and intuitive representation of event data. Unlike conventional methods that primarily focus on sequential event logs, the OCED model captures object interactions, dependencies, and contextual attributes, enabling process miners to gain a more holistic understanding of process dynamics. This shift not only improves analytical depth but also simplifies the interpretation of complex process behaviors.
Furthermore, the implementation of OCEDO on the BPIC 2013 dataset illustrated its practical utility in uncovering hidden patterns, improving traceability, and supporting more informed decision-making. By adopting an object-centric perspective, organizations can move beyond simplistic event logs and embrace a richer, more expressive framework for process analysis. Future research could explore the scalability of this approach, its integration with other process mining techniques, and its application across diverse industries. Furthermore, empirical evaluations, benchmarking, and validation on additional real-world datasets are planned.
%
%
\bibliographystyle{splncs04}
\bibliography{ref}
\end{document}